# Variations of the Hydrogen Bonding and of the Hydrogen Bonded Network in Ethanol-Water Mixtures on Cooling

*Szilvia Pothoczki[1], László Pusztai[1,2] and Imre Bakó[3]*

[1]Wigner Research Centre for Physics, Hungarian Academy of Sciences, H-1121 Budapest, Konkoly Thege M. út 29-33., Hungary

[2]International Research Organization for Advanced Science and Technology (IROAST), Kumamoto University, 2-39-1 Kurokami, Chuo-ku, Kumamoto 860-8555, Japan

[3]Research Centre for Natural Sciences, Hungarian Academy of Sciences, H-1117 Budapest, Magyar tudósok körútja 2., Hungary

**Abstract**

Extensive molecular dynamics computer simulations have been conducted for ethanol-water liquid mixtures in the water-rich side of the composition range, with 10, 20 and 30 mol % of the alcohol, at temperatures between room temperature and the experimental freezing point of the given mixture. All-atom type (OPLS) interatomic potentials have been assumed for ethanol, in combination with two kinds of rigid water models (SPC/E and TIP4P/2005). Both combinations have provided excellent reproductions of the experimental X-ray total structure factors at each temperature; this provided a strong basis for further structural analyses. Beyond partial radial distribution functions, various descriptors of hydrogen bonded assemblies, as well as of the hydrogen bonded network have been determined from the simulated particle configurations. A clear tendency was observed towards that an increasing proportion of water molecules participate in hydrogen bonding with exactly 2 donor- and 2 acceptor sites as temperature decreases. Concerning larger assemblies held together by hydrogen bonding, the main focus was put on the properties of cyclic entities: it was found that, similarly to methanol-water mixtures, the number of hydrogen bonded rings has increased with lowering temperature. However, for ethanol-water mixtures the dominance of not the six-, but of the five-fold rings could be observed.

# 1. Introduction

Mixtures of water and alcohols show significant deviations relative to the ideal solutions in terms of various thermodynamic properties, e.g., self-diffusion coefficient, shear viscosity, excess volume, and excess enthalpy, compressibility, sound attenuation coefficient.[1-14] In most cases, such non-ideal behavior is most pronounced (showing minima or maxima) in the low alcohol concentration region.[1-4,6-13] The perturbation of the local and global structure of the hydrogen-bond (HB) network is now commonly accepted to be the reason behind these anomalous properties.[14] Despite considerable effort in both experiment and theory, significant disagreement remains regarding the microscopic details of this effect.[14,16-18] Different models for the structure of these liquid mixtures have been proposed to address this behavior, including the enhancement of the water hydrogen bonding network around the alcohol hydrophobic groups[14-22] and microscopic immiscibility or clustering.[23-34]

The response of hydrogen-bonded networks to changing thermodynamic variables (and also, constituents) is a key issue in many areas of natural sciences, particularly in chemistry and biology: this has a great deal to do with how living organisms react to decreasing/increasing temperatures. Yet, there is a lack of experimental data on the structure of even simple hydrogen bonded systems, like that of alcohol-water liquid mixtures, as a function of temperature. One of the few available experimental works is the systematic X-ray diffraction investigation of Takamuku et al.,[35] who measured the total structure factor of liquid mixtures of methanol, ethanol and 2-propanol with water. They restricted their work to the water-rich side of the composition range, up to 40 mol% of alcohol, and to temperatures lower than room temperature.

Quite recently, a short account on the structure of methanol-water liquid mixtures has appeared[36] that suggested that the number of 6-fold hydrogen bonded rings had drastically increased when lowering the temperature to near the freezing point. This finding was based on molecular dynamics computer simulation results that closely matched the corresponding X-ray data of Takamuku et al.[35] This promising development has prompted us to move to the other alcohols of the series and perform more extensive computer simulations, with the hope of gaining a better understanding of the behavior of such basic hydrogen bonded liquids.

Here we consider ethanol-water mixtures, with ethanol contents of 10, 20 and 30 mol%, at temperatures between ambient and the freezing point of the actual mixture. We conduct molecular dynamics simulations for each mixture and at each experimental temperature for which structural data are provided by Takamuku et al.[35] The validity of the simulated models is assessed by comparing measured and calculated total structure factors. Particle configurations (sets of atomic coordinates) from simulations are then subjected to detailed geometrical analyses, focusing on hydrogen bonding properties and the network of hydrogen bonds. Observable trends on varying the temperature are revealed in terms of quite a few characteristics, such as size distributions of cyclic entities, etc…

## 2. Methodology

*2.1 Diffraction experiments*

Direct structural information on the structure of liquids can be obtained by the X-ray and neutron diffraction techniques.[37] Results obtained by these methods are usually complementary: X-ray diffraction is sensitive to the electron density and it is proportional to the number of electrons, while neutron diffraction is not. The latter is sensitive to the positions of the nuclei and the scattering 'strengths' of D ($^2$H), C and O (0.667, 0.664 and 0.583 fm) are very similar. These experimental results, coupled with simulation analyses, help us to understand more deeply the local structure of a liquid mixture. Comparison with experimental structure factor (or with composite radial distribution function) is of primary importance for validating results of computer simulation methods (see, e.g., refs 36 and 38).

The part of the total structure function that provides a description for the liquid structure may be calculated from the partial rdf's according to the equation

$$H(k) = \sum_{\alpha \geq \beta} \sum \frac{(2-\delta_{\alpha\beta})x_\alpha x_\beta f_\alpha f_\beta h_{\alpha\beta}(k)}{M(k)} \quad (1)$$

Here, $f_\alpha$ is the scattering length or scattering factor of the α-type atom (which depend on $k$ in the case of X-ray diffraction, and is constant in the case of neutron diffraction) and $x_\alpha$ is the corresponding mole fraction. $M(k) = (\sum_{\alpha=1}^{n} x_\alpha f_\alpha)^2$, $h_{\alpha\beta}(k)$, the partial structure factor, is defined from the partial rdf's, $g_{\alpha\beta}(r)$, according to the following equation:

$$h_{\alpha\beta}(k) = 4\pi\rho \int_0^{r_{max}} r^2 (g_{\alpha\beta}(r)-1) \frac{\sin(kr)}{kr} dr \quad (2)$$

where ρ is the atomic number density of the liquid.

*2.2. Computational details*

Classical molecular dynamics simulations in the NPT and NVT ensembles have been performed, as a function of (decreasing) temperature, on ethanol-water liquid mixtures at $x_e$ = 0.1, 0.2 and 0.3 ethanol mole fraction using the GROMACS software,[39] version 5.1.1. Ethanol molecules were modelled using the OPLS-AA[40] force field with the same parameters (atom types and charges) as in ref. 38. Water molecules were represented by both the SPC/E[41] and TIP4P/2005[42] models. The Lorentz-Berthelot combination rules were used to calculate the cross terms of the potential functions. Bondlengths were kept fixed by the LINCS algorithm[43] for ethanol molecules, while the geometry of water molecules was maintained by the SETTLE[44] algorithm. The Newtonian equations of motions were integrated via the leapfrog algorithm, using a time step of 1 fs. The particle-mesh Ewald algorithm was used for handling the long range electrostatic forces and potentials[45,46]. The cut-off radius for non-bonded interactions was set to 1.1 nm. The applied temperatures, box lengths with the corresponding number densities and densities are given in Supplementary Material.

Initially, an energy minimization procedure was performed for each composition, at room temperature, using the steepest descent method. This was followed by a 1 ns equilibration run and then, a 5 ns production run in the NVT ensemble; the temperature (298 K) was controlled by a Berendsen thermostat[47] with temperature coupling time constants set to 0.1 ps and 0.5 ps, respectively. At the lower temperature values, 2ns NPT simulations were utilized with Berendsen thermostat and Berendsen barostat;[47] both coupling time constants were 0.1 ps. Then, in order to achieve full equilibration, long (10 ns) NPT simulations were conducted, where the temperature was maintained via the Nose-Hoover[48,49] thermostat with $\tau_T$=1.0 ps, while the pressure was held at 1 bar via the Parrinello-Rahman[50] barostat using $\tau_T$=3.0 ps. To complete the simulation process, at each experimental temperature, a 5 ns production NVT run was performed using the densities obtained from NPT simulations described, using the Nose-Hoover[48,49] thermostat with $\tau_T$=0.5 ps,

The total scattering structure factors and the radial distribution functions were calculated based on the last 500 frames of the NVT production runs.

Three independent systems, with different compositions, were modelled according to the available diffraction data: 336, 576 and 756 ethanol molecules and 3024, 2304 and 1764 water molecules provided solutions with ethanol mole fractions of 0.1, 0.2 and 0.3, respectively. The box lengths corresponded to experimental densities.

## 3. Results

### 3.1. Total scattering structure factors

F(Q) from experiments and simulations for the investigated systems are presented in Figure 1 for $x_e$=0.2 ethanol-water mixture as a function of temperature. Based on mere visual inspection, the excellent agreements shown might even be called as (at least semi-) 'quantitative'.

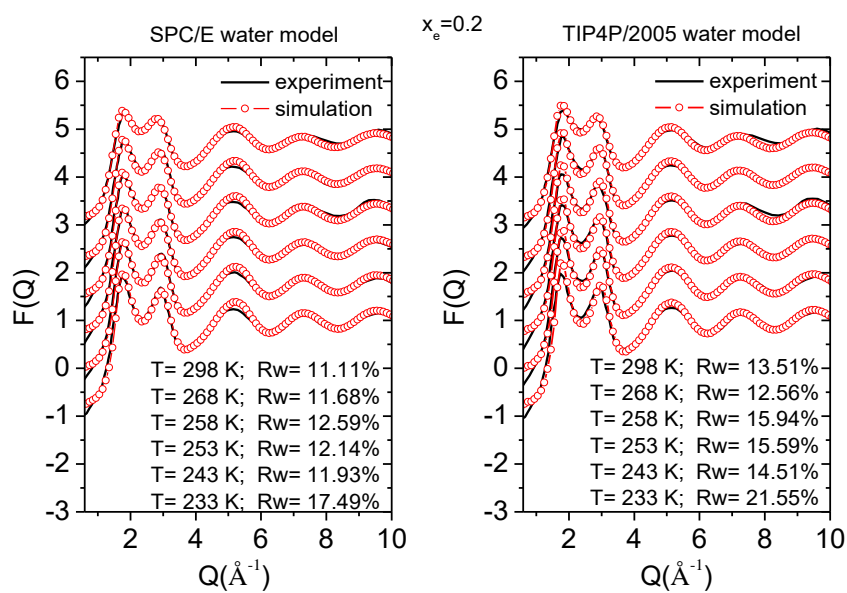

Figure 1 Total scattering structure factors for the mixture with 20 mol % ethanol, as a function of temperature. Solid line: X-ray diffraction data; symbols: molecular dynamics simulations. Left panel: SPC/E water model; Right panel: TIP4P/2005 water model.

Recently Gereben et al.[38] characterized the performance of different water models by the difference, $R_w$ (goodness-of-fit), between MD simulated (averaged over many time frames) and experimental structure factors for ethanol/water mixtures at 298 K. At higher water contents, the best agreement was achieved by using the TIP4P/2005 force field, but the SPC/E model also provided a reasonable agreement.

Here also the same measure was chosen for making the difference between the two water models applied. The calculated goodness-of-fits are presented in Table 1. We can conclude that at low temperature the SPC/E water model provides the better agreement in general. At room temperature, both the SPC/E and TIP4P/2005 model gave reasonably good results.

|       | $x_e$=0.1 | | $x_e$=0.2 | | $x_e$=0.3 | |
|---|---|---|---|---|---|---|
|       | $R_w$ (%) SPC/E | $R_w$ (%) TIP4P/2005 | $R_w$ (%) SPC/E | $R_w$ (%) TIP4P/2005 | $R_w$ (%) SPC/E | $R_w$ (%) TIP4P/2005 |
| 298 K | 10.88 | 9.76  | 11.11 | 13.51 | 14.06 | 13.02 |
| 268 K | 9.88  | 9.73  | 11.68 | 12.56 | 14.91 | 13.27 |
| 258 K | 13.64 | 14.79 | 12.59 | 15.94 |       |       |
| 253 K | 15.15 | 22.71 | 12.14 | 15.59 | 16.97 | 17.37 |
| 243 K |       |       | 11.93 | 14.51 |       |       |
| 238 K |       |       |       |       | 15.66 | 14.08 |
| 233 K |       |       | 17.49 | 21.55 |       |       |

Table 1 Comparison of the Goodness-of-fits for All Models of Ethanol-Water Mixtures.

*3.2 Partial radial distribution functions and coordination numbers*

The structure of (also, H-bonded) liquids can be described in terms of the partial radial distribution functions, $g_{\alpha\beta}(r)$, at the two-particle level. The most interesting and informative $g_{\alpha\beta}(r)$ functions for understanding the structure of the hydrogen bonding interaction in our systems are the following: $O_{eth}O_{eth}$, $O_{eth}O_{wat}$, $O_{wat}O_{wat}$ (Figure 2), $O_{eth}OH_{eth}$, $O_{eth}H_{wat}$, $O_{wat}OH_{eth}$, $O_{wat}H_{wat}$ (Figure 3; 'OH$_{eth}$' denotes the hydroxyl H of ethanol molecules). Additionally, the $C_{eth}C_{eth}$ and $C_{eth}O_{eth}$ prdf-s can be found in Figure 2 that provide information about the carbon backbone of ethanol molecules. These figures show results at 6 different temperatures for the $x_e$=0.2 solution, using both the SPC/E and TIP4P/2005 water models. The same results for the other two compositions are presented in Supp. Material. The $g_{\alpha\beta}(r)$ functions at 298 K are in agreement with those from earlier computer simulations (see, e.g., ref. 38). Also note that PRDF-s from the SPCE/E water model resemble very closely to those from the TIP4P/2005 potential; for this reason, in the later sections results only from one of them, the SPC/E potential, will be reported.

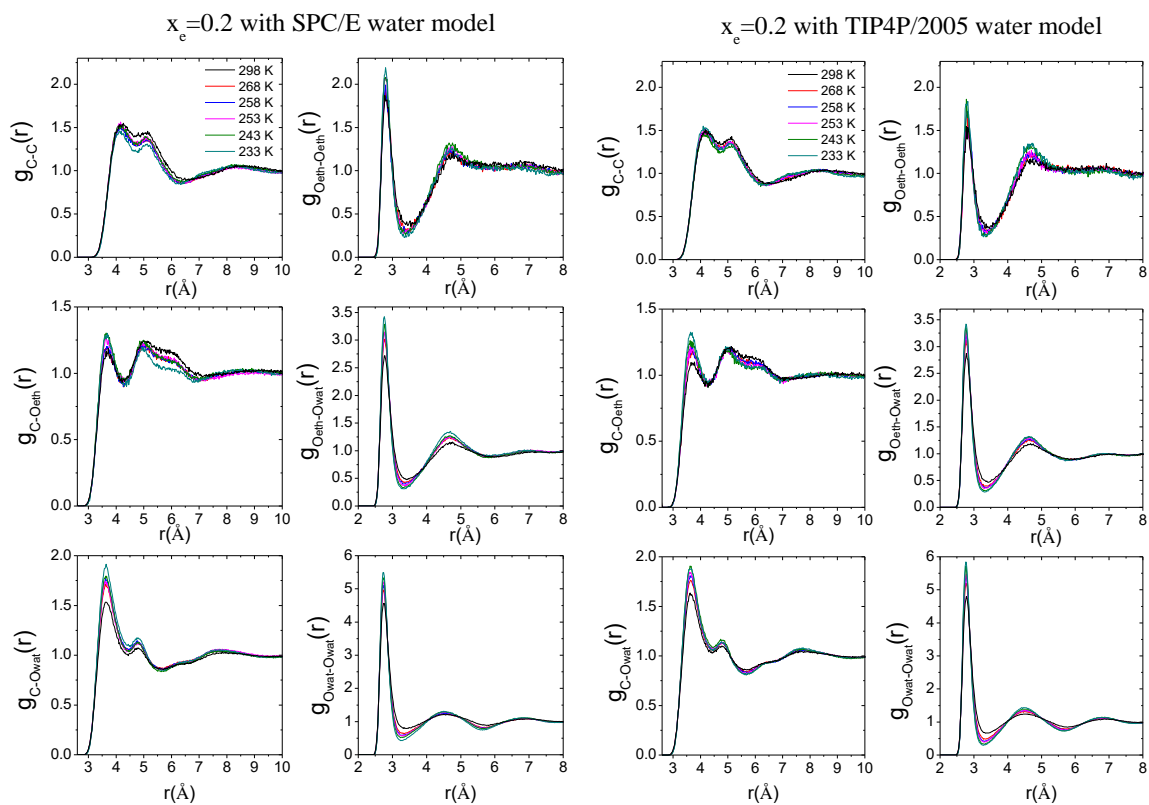

Figure 2 Heavy-atom related partial radial distribution functions for the mixture with 20 mol% ethanol, as a function of temperature. Left panel: SPC/E water model; Right panel: TIP4P/2005 water model.

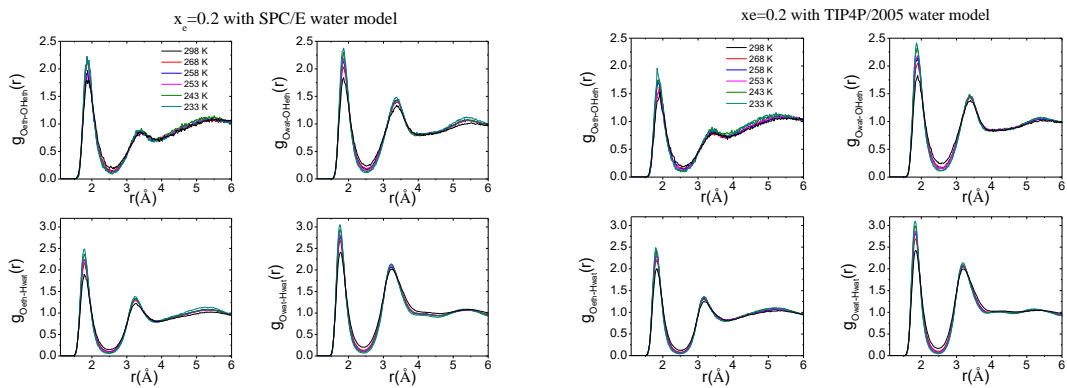

Figure 3 H-bond related partial radial distribution functions for the mixture with 20 mol % ethanol, as a function of temperature. Left panel: SPC/E water model; Right panel: TIP4P/2005 water model.

The characteristic quantities of these $g_{\alpha\beta}(r)$ (peak positions, peak heights, coordination numbers calculated up to the first minimum) can be found in Supp. Mat. It can be concluded from our data that structural features (at the 2-body level) become more pronounced when temperature decreases. The main effect of decreasing temperature is that the first peak heights become higher; while the positions of maxima and minima change only according to the slightly changing density. For both water models considered the closest $O_{wat}H_{wat}$, $O_{eth}H_{wat}$ and $O_{wat}OH_{eth}$ distances are considerable shorter than the shortest $O_{eth}OH_{eth}$ one. Coordination numbers calculated up to the first minima of the PRDF-s do not change significantly with decreasing temperature. As it was already known previously, at room temperature the $O_{wat}O_{wat}$ coordination numbers are decreasing, while the $O_{eth}O_{eth}$ and $O_{eth}O_{wat}$ ones are increasing with increasing ethanol concentration.

*3.3. Hydrogen-bonding*

The physical and chemical properties of hydrogen bond network forming liquids reflect the extended intermolecular structures that result from hydrogen bond formation. One of the constituents of these mixtures (water) is a "random" three-dimensional hydrogen bonded network system where individual water molecules form tetrahedral hydrogen bonded surroundings with their neighboring water molecules. Analysis of hydrogen bond statistics can give us information about the average local structure of a liquid in terms of a distribution of the number of molecules in positions forming hydrogen bonds with the central one.

In MD simulations the interaction energy and intermolecular variables (distances, angles) corresponding to an H-bond are changing continuously. Consequently, criteria for deciding that an H-bond is established are somewhat arbitrary. A widely accepted definition of molecules forming a hydrogen bond with the one in the center of the coordinate system is usually based on energetic or geometric criteria (see, e.g., ref. 51).

Like in our earlier study of methanol-water mixtures,[36] both purely geometric, as well as energetic definitions have been applied here for identifying hydrogen bonds. In the present study, the energetic definition of H-bonds was used as follows: two molecules are considered to be hydrogen bonded to each other if they are found at a distance $r(O\cdots H) < 2.5$ Å, and the interaction energy is smaller than $-12$ kJ/mol. Note that purely geometric definitions do not necessarily have direct implications on the energetics of H bonding. We have therefore carefully tested how our conclusions depend on the applied definition. As usage of both hydrogen-bond definitions led to the same conclusions, only results obtained from the OPLS-AA-ethanol -- SPC/E-water model using the energetic definition will be presented.

In Figure 4 the average hydrogen bond numbers, as a function of temperature, for the investigated three systems are presented. These 'cumulative' values have been decomposed to various terms in order to obtain specific information on the water-water and ethanol-water H-bonds. It can be concluded from the corresponding curves that the average number of hydrogen bonds per molecule increases by about 10 % as temperature decreases. This statement is valid in the case of the average number of H-bonded water molecules around water and ethanol, too. The change in terms of the average number of H-bonded ethanol molecules around ethanol is not significant (not shown here).

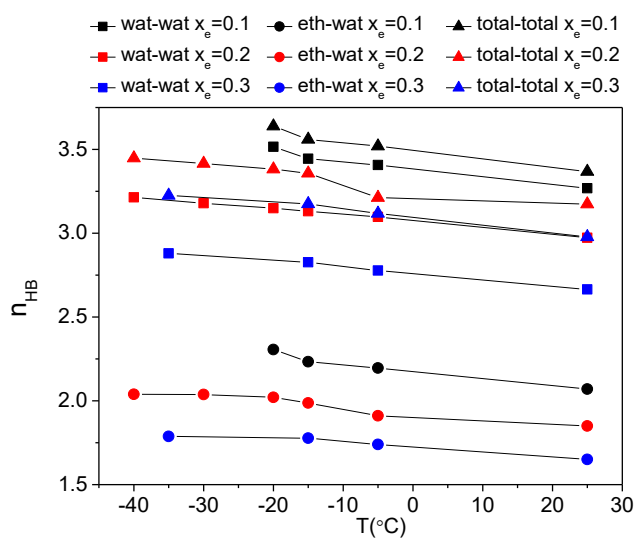

Figure 4 Average number of hydrogen bonds/molecule as a function of composition and temperature. Black symbols: $x_e$= 0.1; red symbols: $x_e$= 0.2; blue symbols: $x_e$=0.3. Squares: water around water; circles: water around ethanol; triangles: sum of all contributions.

Molecules can be classified based on the number of hydrogen-bonds they take part in as H-acceptors ($n_A$=0, 1, 2 for ethanol and $n_A$=0, 1, 2, 3 for water molecules[52]) and H-donors ($n_D$=0, 1 for ethanol and $n_D$=0, 1 or 2 for water). Molecules are thus tagged as ($n_A$,$n_D$), e.g. (0,0) – no bond, up to (3,2) – fully bonded. On the acceptor side, due to the significantly larger space on this side of the water molecule, even 3 water molecules may approach via H-bonds (3,2).[52] Calculated data are presented in Table 2 and 3.

| ethanol (H-acceptor, H-donor) | | | | |
|---|---|---|---|---|
| T (K) | (1,1) | (2,1) | (1,0) | (2,0) |
| $x_e$=0.1 | | | | |
| 298 | 0.338 | 0.363 | 0.159 | 0.122 |
| 268 | 0.325 | 0.442 | 0.119 | 0.105 |
| 258 | 0.323 | 0.466 | 0.103 | 0.102 |
| 253 | 0.309 | 0.521 | 0.081 | 0.083 |
| $x_e$=0.2 | | | | |
| 298 | 0.412 | 0.306 | 0.171 | 0.092 |
| 268 | 0.412 | 0.331 | 0.131 | 0.081 |
| 258 | 0.413 | 0.389 | 0.114 | 0.076 |
| 253 | 0.417 | 0.399 | 0.106 | 0.071 |
| 243 | 0.413 | 0.420 | 0.096 | 0.066 |
| 233 | 0.428 | 0.427 | 0.085 | 0.057 |
| $x_e$=0.3 | | | | |
| 298 | 0.469 | 0.259 | 0.174 | 0.070 |
| 268 | 0.486 | 0.310 | 0.131 | 0.060 |
| 258 | 0.496 | 0.328 | 0.112 | 0.054 |
| 238 | 0.511 | 0.342 | 0.094 | 0.047 |

Table 2 The Fractions of Ethanol Molecules as H-acceptors and as H-donors in the H-bonds Identified, as a Function of Temperature and Concentration.

| water (H-acceptor, H-donor) | | | | | |
|---|---|---|---|---|---|
| T (ºC) | (2,1) | (1,2) | (2,2) | (1,1) | (3,2) |
| $x_e$=0.1 | | | | | |
| 298 | 0.114 | 0.232 | 0.522 | 0.071 | 0.028 |
| 268 | 0.084 | 0.196 | 0.639 | 0.039 | 0.025 |
| 258 | 0.074 | 0.183 | 0.677 | 0.031 | 0.024 |
| 253 | 0.052 | 0.152 | 0.756 | 0.018 | 0.015 |
| $x_e$=0.2 | | | | | |
| 298 | 0.099 | 0.278 | 0.492 | 0.079 | 0.020 |
| 268 | 0.071 | 0.267 | 0.568 | 0.056 | 0.017 |
| 258 | 0.063 | 0.238 | 0.637 | 0.036 | 0.015 |
| 253 | 0.057 | 0.230 | 0.658 | 0.031 | 0.014 |
| 243 | 0.048 | 0.210 | 0.687 | 0.023 | 0.012 |
| 233 | 0.040 | 0.204 | 0.721 | 0.019 | 0.011 |
| $x_e$=0.3 | | | | | |
| 298 | 0.086 | 0.324 | 0.454 | 0.087 | 0.015 |
| 268 | 0.061 | 0.301 | 0.556 | 0.050 | 0.013 |
| 258 | 0.049 | 0.285 | 0.607 | 0.036 | 0.011 |
| 238 | 0.038 | 0.267 | 0.654 | 0.025 | 0.009 |

Table 3 The Fractions of Water Molecules as H-acceptors and as H-donors in the H-bonds Identified, as a Function of Temperature and Concentration.

A clear preference for four-fold H-bonded water molecules with two acceptors and two donors, denoted by (2, 2), has been observed at each concentration. For all concentrations the fraction of 4 H-bonded water molecules increases as temperature decreases. In addition, the fraction of 5 bonded water molecules is decreasing with temperature. There is a well-defined asymmetry between (1,2) and (2,1) type water molecules, in terms of their populations, and this difference is more pronounced at lower temperatures. Concerning the other constituent, the fraction of (2,1) ethanol molecules is increasing as the temperature is decreasing. In summary, therefore, we can conclude that for all concentrations the 'ideal' H-bonded coordination of water and ethanol molecules becomes as populated as possible when the temperature decreases.

*3.4 Hydrogen bonded assemblies ('clusters') and properties of the hydrogen bonded network*

In order to understand the nature and evolution of H-bonded networks present in such binary systems as a function of temperature, it is essential to observe the propensity of cluster formation. Here, two molecules are regarded as belonging to the same cluster if they are connected by a chain of hydrogen bonds. The size of a cluster, $n_c$, is characterized by the number of the molecules belonging to it. The cluster size is obviously dependent on the connectivity within the liquid, which, in turn, is related to the number of hydrogen bonds.

Percolation can be monitored by comparing the calculated cluster size distribution function for the present systems with that obtained for random percolation on a 3D cubic lattice, in which $P(n_c) = n_c^{-2.19}$ ($n_c$ is the number of molecules in a given cluster), cf. ref. 53 (and references therein). In percolating systems the cluster size distribution exceeds this predicted function at large cluster size values. Figure 5 presents the hydrogen-bonded cluster size distribution in water-ethanol mixtures at various concentrations. In each mixture the molecules percolate throughout the entire system. Figure 6 shows the hydrogen-bonded cluster size distributions at the concentration of $x_e=0.3$ for the water-water and ethanol-ethanol (sub)networks. At this concentration water molecules are percolated throughout the system, whereas the ethanol subsystem consists of small (<10 ethanol molecules) isolated clusters. These findings are valid for the other two mixtures, too.

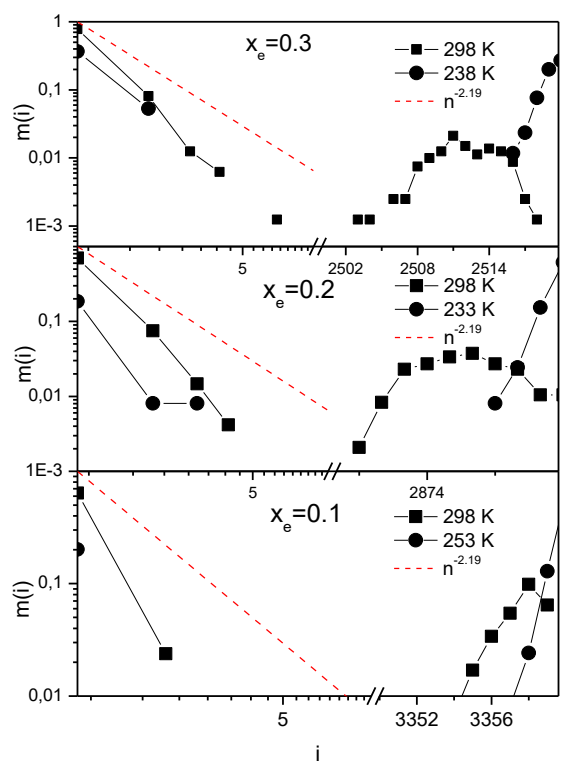

Figure 5 Size distribution of hydrogen-bonded clusters in water-ethanol mixtures at various concentrations and two temperatures. The dashed lines represent the percolation limit that may be estimated from random percolation on a 3d lattice.

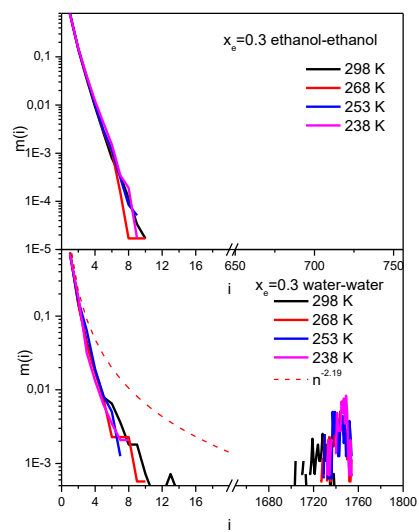

Figure 6 Size distributions of hydrogen-bonded ethanol (upper panel) and water (lower panel) clusters in water-ethanol mixtures at the concentration of $x_e=0.3$, as a function of temperature. The dashed line represents the percolation limit that may be estimated from random percolation on a 3d lattice.

The structure of complex networks can be characterized by their topological properties. Typical hydrogen bond network topologies for all the molecules, as well as for the water-water and ethanol-ethanol subsystems, at the composition of $x_e=0.3$, are shown in Figure 7. In these figures we can detect two different types of structural units, namely cyclic structures and chains, or branched chain-like aggregations. Clusters can contain chains "closed into themselves", i.e., the last molecule along a hydrogen-bonded chain can form a bond with the first one, forming a cycle that we can call primitive cycle or ring. It can be clearly seen in these figures that in network topologies for all molecules, as well as for the water subsystem, contain several cyclic entities. On the other hand, in the ethanol subsystem, ethanol molecules form only short chain like structures.

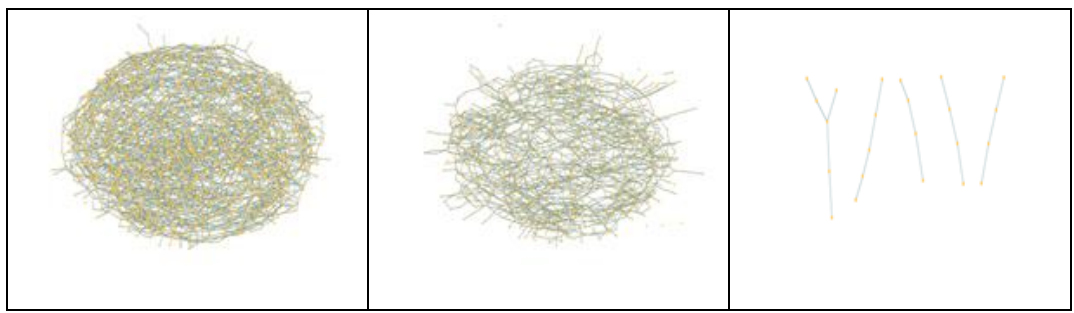

**Figure 7** Typical hydrogen bond network topologies for all the molecules (left panel), and for the water-water (middle panel) and ethanol-ethanol (right panel) subsystems, at the composition of $x_e$=0.3

Molecules can be said to participate in a given cyclic entity if there is a minimum length path consisting of a series of hydrogen bonds through which one can get back to the original molecule. We can define the size of a cyclic entity as the shortest path through the hydrogen bonds. To estimate the distribution of such cyclic entities, the ring search algorithm developed by Chiaia et al.[54] was used in this work; the same method has already been used in the investigation of the hydrogen-bonded network topology in liquid formamide[55] and in water-methanol mixtures[36], as well. For water-methanol mixtures at low temperature, we found that the overall number of cycles becomes progressively larger, especially for the 6 membered cycles, as temperature decreases.

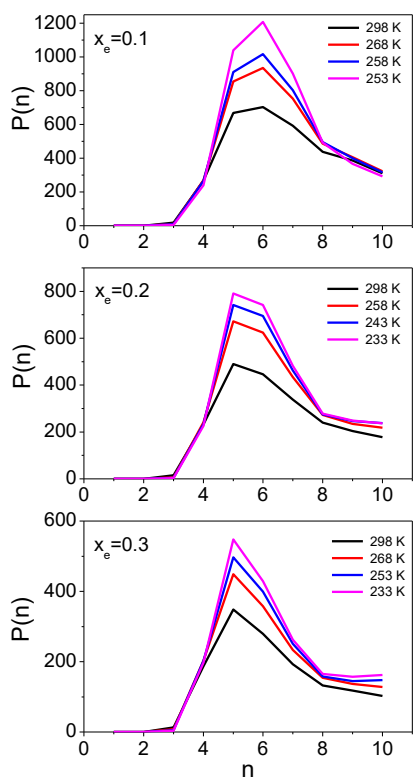

Figure 8 Ring size distributions for the three mixtures (upper panel: $x_e = 0.1$; middle panel: $x_e = 0.2$; lower panel: $x_e = 0.3$), as a function of temperature. H-bonds have been identified on the basis of the energetic criterion.

Ring size distributions (not normalized) for the three mixtures (upper panel: $x_e = 0.1$; middle panel: $x_e = 0.2$; lower panel: $x_e = 0.3$), as a function of temperature, are presented in Fig. 8. Similarly to the methanol-water case[36], the number of cyclic entities has increased enormously as the temperature dropped to close to the freezing point of the given mixture. The most abundant is the six membered ring at the lowest ethanol concentration ($x_e = 0.1$). On the other hand, at $x_e = 0.2$ and 0.3 the most probable cyclic entities contain 5 (water or ethanol) molecules; this is a marked difference in comparison with methanol-water mixtures. An explanation may be that as the ratio of ethanol molecules grows, the number of ethanol molecules in the ring structure, that need larger volumes if included, also grows (see below).

The sheer size of the hydrophobic ethyl group seems to be sufficient for forcing the H-bonded ring to close sooner (i.e., with fewer molecules in the ring), by not allowing more molecules to approach the ring being formed, than when only water (or methanol, which are smaller than ethanol) molecules form the cyclic entity.

Note that any given molecule can belong to more than one ring. We calculated the number of water and ethanol molecules that participate in a cyclic entity with a given size ($m_c$ denotes the size of the cycle). We showed earlier that if in a two component mixture the constituent molecules are similar in this sense then the ratio of these two numbers, calculated for the two components, should be very close to the molar ratio of the components for each $m_c$, like in water-formamide mixtures (see ref. 54). In contrast, if the behavior, the H-bonding character, of the constituents is different (e.g. water-methanol mixtures), then the ratio may be very far from the molar ratio.

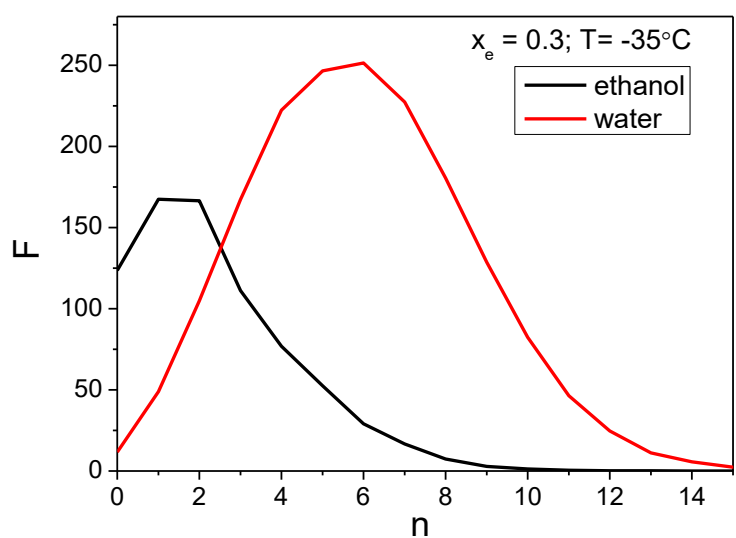

Figure 9 Distribution of the number of cycles that go through a given molecule in the mixture with $x_e$ =0.3, and at a temperature of 238 K (n: number of cycles; F: the number of ethanol (black solid) and water (red solid) molecules that are parts of exactly 'n' cycles).

The distribution of the quantities for water and ethanol in $x_e=0.3$ is presented in Fig 9. In this mixture one water molecule participates in the formation of more than even 10 rings (the most probable value is 7), but ethanol molecules belong to maximum 6 rings (the most probable value is 3). These quantities were decomposed according to the average H-bond number of ethanol or water and presented in Fig. 10; these results are the same for every concentration. We can conclude from these figures that this type of different H-bonding character mainly arises from the presence of 4-bonded water molecules. On the basis of the analysis of the ratio of the number of rings in which water and ethanol molecules participate, it is revealed that the composition of cyclic entities in ethanol-water mixtures is unbalanced and that this missing balance is most pronounced at low ethanol concentration.

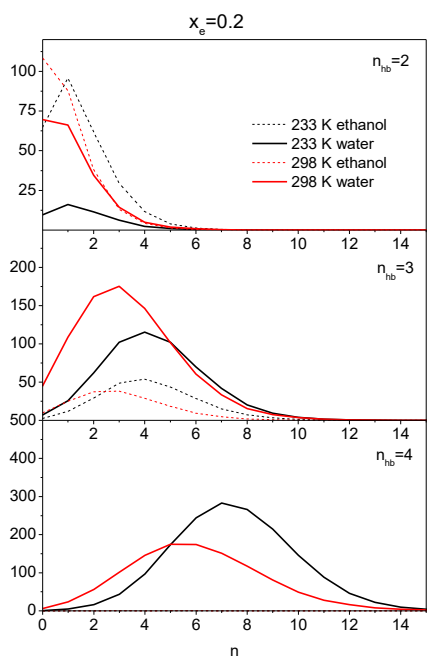

Figure 10 Same as Figure 9, but for $x_e=0.2$ and decomposed according to the number of H-bonds/molecule (i.e., by the 'H-bonding state' of molecules). Upper panel: $n_{hb}=2$; middle panel: $n_{hb}=3$; lower panel: $n_{hb}=4$ (for this, results only for water molecules exist, since there are no ethanol molecules with 4 H-bonds). Black lines: 233 K; red lines: 298 K.

Further, we calculated the average number of ethanol molecules incorporated in cyclic structures of certain sizes. If the H-bonding character of water and ethanol molecules was identical at this level than this number should be approximately equal to, where $m_c$ is the size of the ring and $x_e$ is the molar fraction of ethanol. The corresponding results are presented in Fig 11; clearly, the calculated values at every concentration and temperature are significantly smaller than $m_c*x_e$. The deviation from the balanced behavior is more pronounced in smaller rings. This kind of imbalance decreases somewhat with decreasing temperature, especially for larger ring sizes.

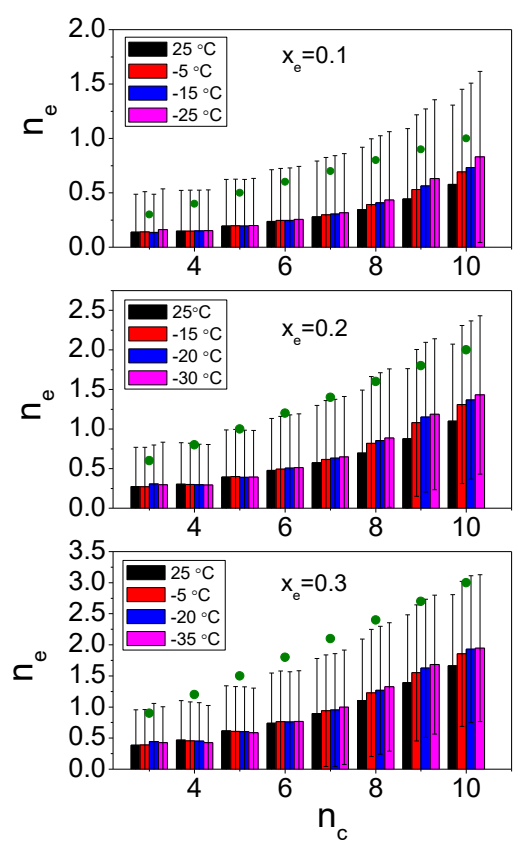

Figure 11 Average number of ethanol molecules incorporated in cyclic structures of certain sizes, as a function of temperature and composition (upper panel: $x_e$ =0.1; middle panel: $x_e$ =0.2; lower panel: $x_e$ =0.3). Green dots indicate the number of ethanol molecules in a given sized ring that would be if the participation of ethanol molecules would be proportional to their molar fraction. (Note also the large error bars.)

**4. Summary, conclusions and outlook**

Extensive molecular dynamics computer simulation studies have been performed on ethanol-water mixtures, as a function of decreasing temperature, on solutions containing 10, 20 and 30 molar % of ethanol ($x_e$ = 0.1, 0.2 and 0.3). Interactions of alcohol molecules have been represented by the all-atom type OPLS-AA potentials, whereas two different models, SPC/E and TIP4P/2005, have been tested for water. Quantitative agreement with total structure factors, measured by X-ray diffraction, has been exploited as validation of simulation results. The following findings are thought to be worthy of emphasizing as major conclusions:

(1) The two water potentials were able to reproduce the measured total structure factors roughly equally well; the level of agreement was nearly quantitative at each temperature and for each composition. For further analyses, the SPC/E potential was selected.

(2) A clear tendency towards '2 donor, 2 acceptor' H-bonded sites in water molecules with decreasing temperature could be detected.

(3) Apart from the mixture with 10 mol % ethanol, the dominance of 5-fold rings could be observed, whose ratio has systematically increased with lowering the temperature. This is in contrast with the case of methanol-water mixtures, for which always 6-fold H-bonded rings were the most abundant. This difference may be understood by taking the size difference between methanol and ethanol molecules into account.

As far as further studies are concerned, given the lack of measured data and the novel findings in conjunction with the temperature dependent structure of alcohol-water mixtures, computer simulations for mixtures with isopropanol, as well as X-ray and neutron diffraction experiments over a wider composition range, are underway.

**Acknowledgements**

The authors are grateful to the National Research, Development and Innovation Office (NRDIO (NKFIH), Hungary) for financial support via grants Nos. SNN 116198 and 124885. We thank Dr. Laszlo Temleitner (Wigner RCP, Hungary) for his assistance concerning experimental data.

**Supporting Information**

Additional details of the performed MD simulations (Table S1); comparison of the measured and the calculated (from MD simulations) total scattering structure factors as a function of temperature for the mixture with 10 and 30 mol % ethanol (Figure S1-S2); total scattering structure factors for neutron diffraction („prediction") as a function of temperature in the case of the mixture with 10, 20 and 30 mol % ethanol (Figure S3-S5); heavy-atom related partial radial distribution functions as a function of temperature for the mixture with 10 and 30 mol % ethanol (Figure S6-S7); H-bond related partial radial distribution functions as a function of temperature for the mixture with 10 and 30 mol % ethanol (Figure S8-S9); the characteristic quantities of $g_{\alpha\beta}(r)$ for the mixture with 20 mol % ethanol (Table S2); ratios of the numbers of ethanol and water molecules that are parts of exactly 'n' cyclic entities, as a function of temperature (Figure S10).

Supporting Information

for

**Variations of the hydrogen bonding and of the hydrogen bonded network in ethanol-water mixtures on cooling**

*Szilvia Pothoczki[1], László Pusztai[1,2] and Imre Bakó[3]*

[1]Wigner Research Centre for Physics, Hungarian Academy of Sciences, H-1121 Budapest, Konkoly Thege út 29-33., Hungary

[2]International Research Organization for Advanced Science and Technology (IROAST), Kumamoto University, 2-39-1 Kurokami, Chuo-ku, Kumamoto 860-8555, Japan

[3]Research Centre for Natural Sciences, Hungarian Academy of Sciences, H-1117 Budapest, Magyar tudósok körútja 2., Hungary

**Table S1:** The applied temperatures, box lengths, together with the corresponding number densities and densities.

| | | SPC/E | | | TIP4P/2005 | | |
|---|---|---|---|---|---|---|---|
| $x_e$ | T (K) | L (nm) | number density (atom/Å$^3$) | density (g/cm$^3$) | L (nm) | number density (atom/Å$^3$) | density (g/cm$^3$) |
| 0.10 | 298 | 4.6900 | 0.1173 | 1.126 | 4.6900 | 0.1173 | 1.126 |
| 0.10 | 268 | 4.8892 | 0.1035 | 0.994 | 4.9056 | 0.1025 | 0.984 |
| 0.10 | 258 | 4.8850 | 0.1038 | 0.997 | 4.9076 | 0.1023 | 0.983 |
| 0.10 | 253 | 4.8802 | 0.1041 | 0.999 | 4.9069 | 0.1024 | 0.983 |
| 0.20 | 298 | 4.9500 | 0.0997 | 0.932 | 4.9500 | 0.0997 | 0.932 |
| 0.20 | 268 | 4.8889 | 0.1035 | 0.967 | 4.9032 | 0.1026 | 0.959 |
| 0.20 | 258 | 4.8752 | 0.1044 | 0.975 | 4.8935 | 0.1032 | 0.964 |
| 0.20 | 253 | 4.8752 | 0.1044 | 0.975 | 4.8763 | 0.1043 | 0.975 |
| 0.20 | 243 | 4.8560 | 0.1056 | 0.987 | 4.8667 | 0.1049 | 0.980 |
| 0.20 | 233 | 4.8489 | 0.1061 | 0.991 | 4.8585 | 0.1055 | 0.985 |
| 0.30 | 298 | 5.1900 | 0.0865 | 0.791 | 5.1900 | 0.0865 | 0.791 |
| 0.30 | 268 | 4.8903 | 0.1034 | 0.946 | 4.9000 | 0.1028 | 0.940 |
| 0.30 | 253 | 4.8683 | 0.1048 | 0.959 | 4.8791 | 0.1041 | 0.952 |
| 0.30 | 238 | 4.8405 | 0.1066 | 0.975 | 4.8526 | 0.1059 | 0.968 |

**Figure S1:** Total scattering structure factors for the mixture with 10 mol % ethanol as a function of temperature. Solid line: X-ray diffraction data; symbols: molecular dynamics simulations. Left panel: SPC/E water model; Right panel: TIP4P/2005 water model.

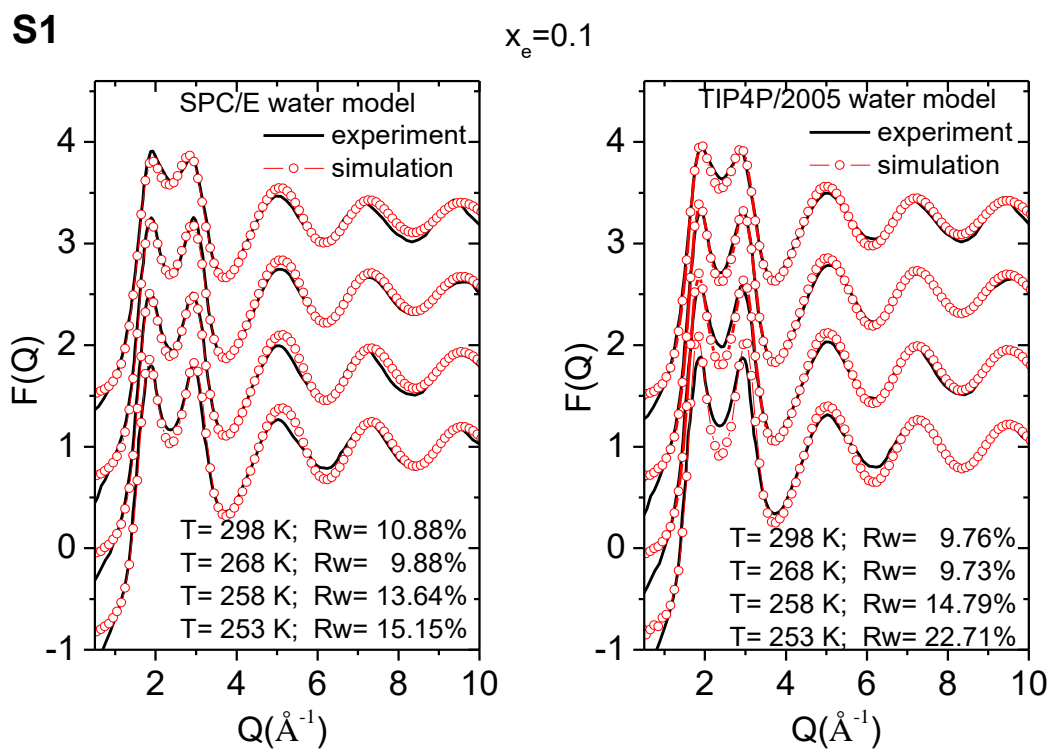

**Figure S2:** Total scattering structure factors for the mixture with 30 mol % ethanol as a function of temperature. Solid line: X-ray diffraction data; symbols: molecular dynamics simulations. Left panel: SPC/E water model; Right panel: TIP4P/2005 water model.

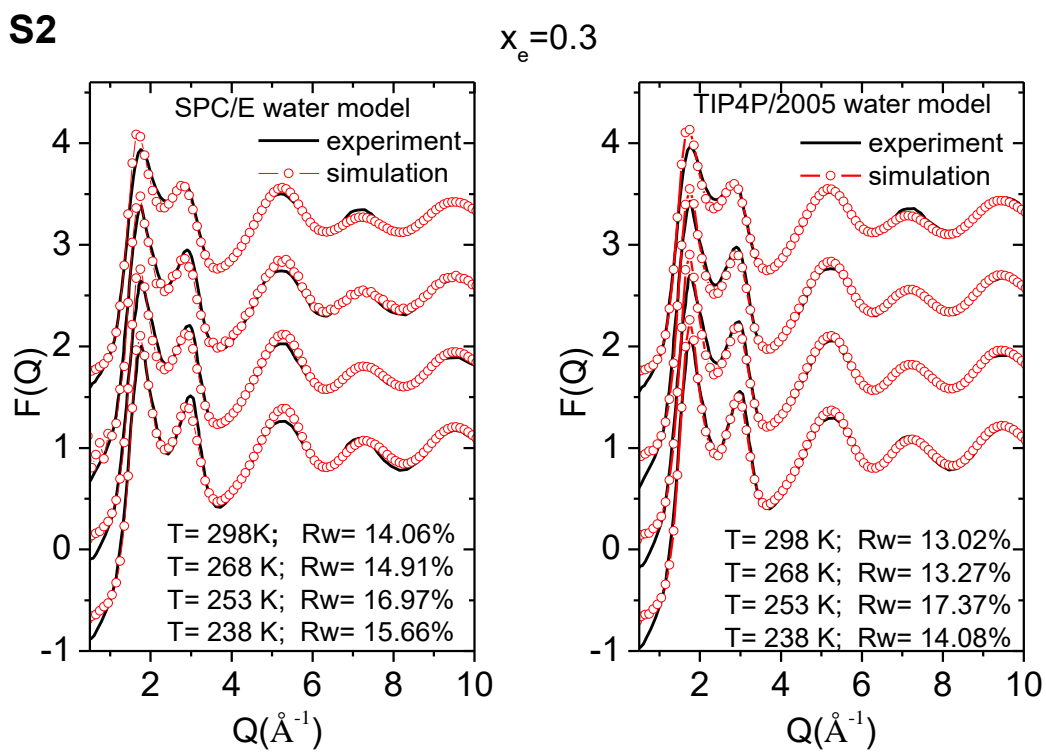

**Figure S3:** Total scattering structure factors for neutron diffraction („prediction") in the case of the mixture with 10 mol % ethanol, as a function of temperature.

**S3**

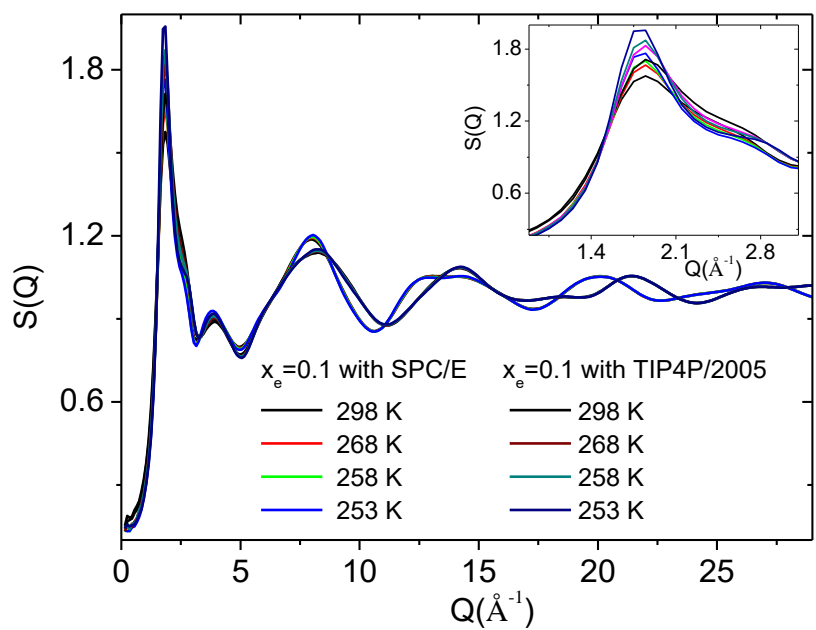

**Figure S4:** Total scattering structure factors for neutron diffraction („prediction") in the case of the mixture with 20 mol % ethanol, as a function of temperature.

**S4**

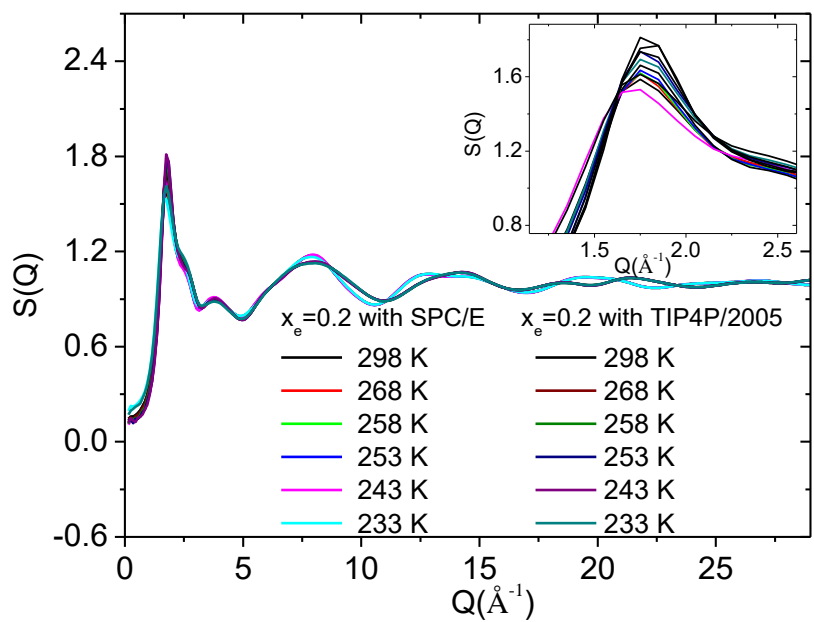

**Figure S5:** Total scattering structure factors for neutron diffraction („prediction") in the case of the mixture with 30 mol % ethanol, as a function of temperature.

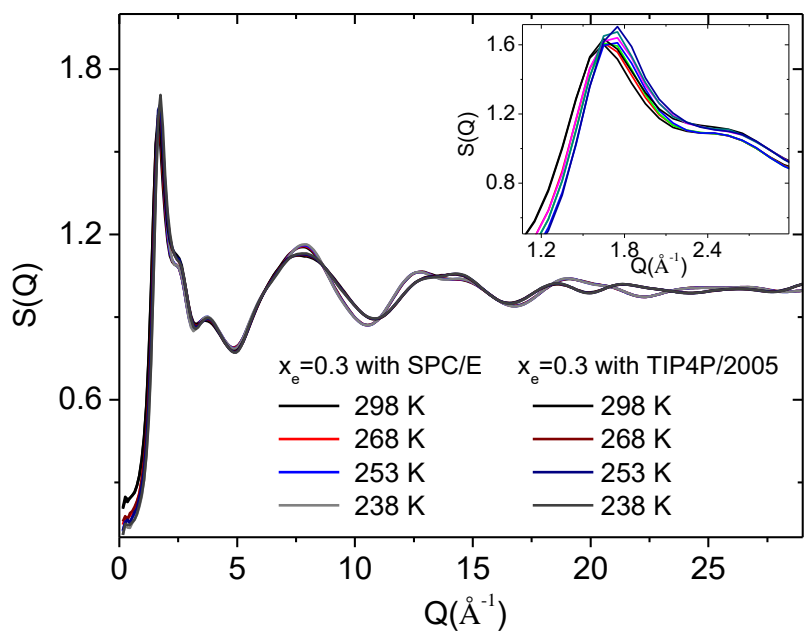

**Figure S6:** Heavy-atom related partial radial distribution functions for the mixture with 10 mol % ethanol, as a function of temperature. a) SPC/E water model; b) TIP4P/2005 water model.

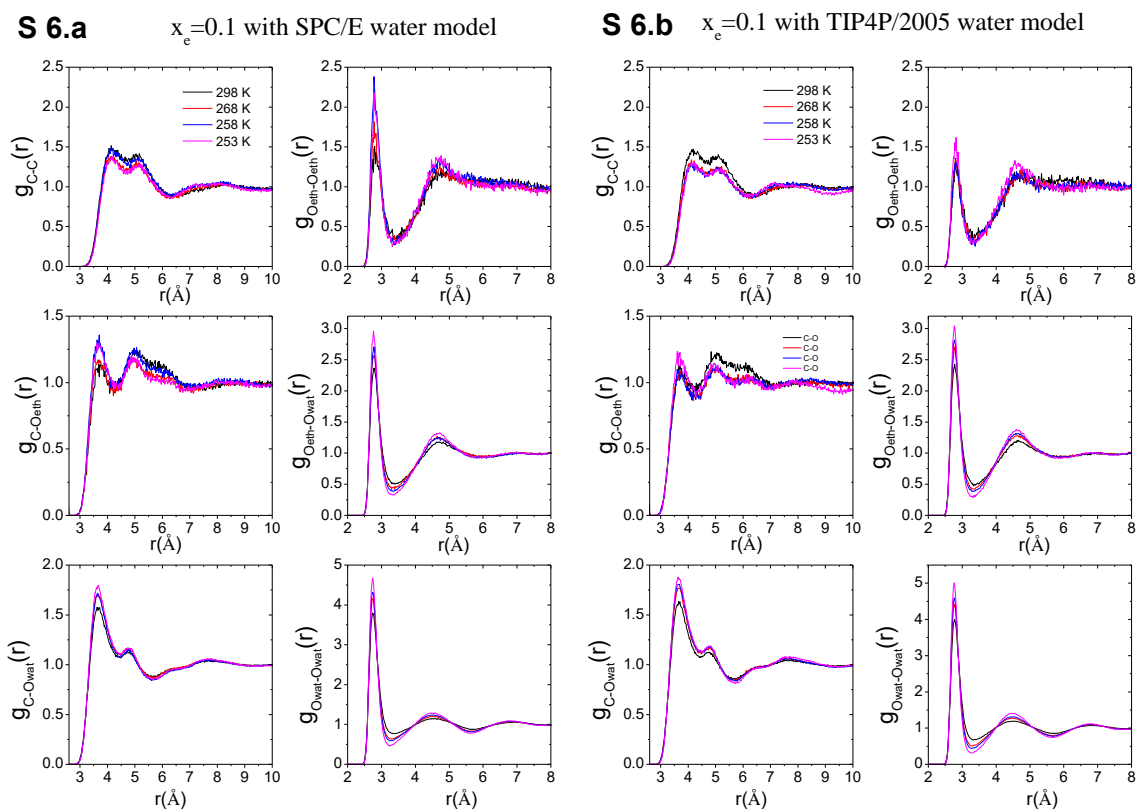

**Figure S7:** Heavy-atom related partial radial distribution functions for the mixture with 30 mol % ethanol, as a function of temperature. a) SPC/E water model; b) TIP4P/2005 water model.

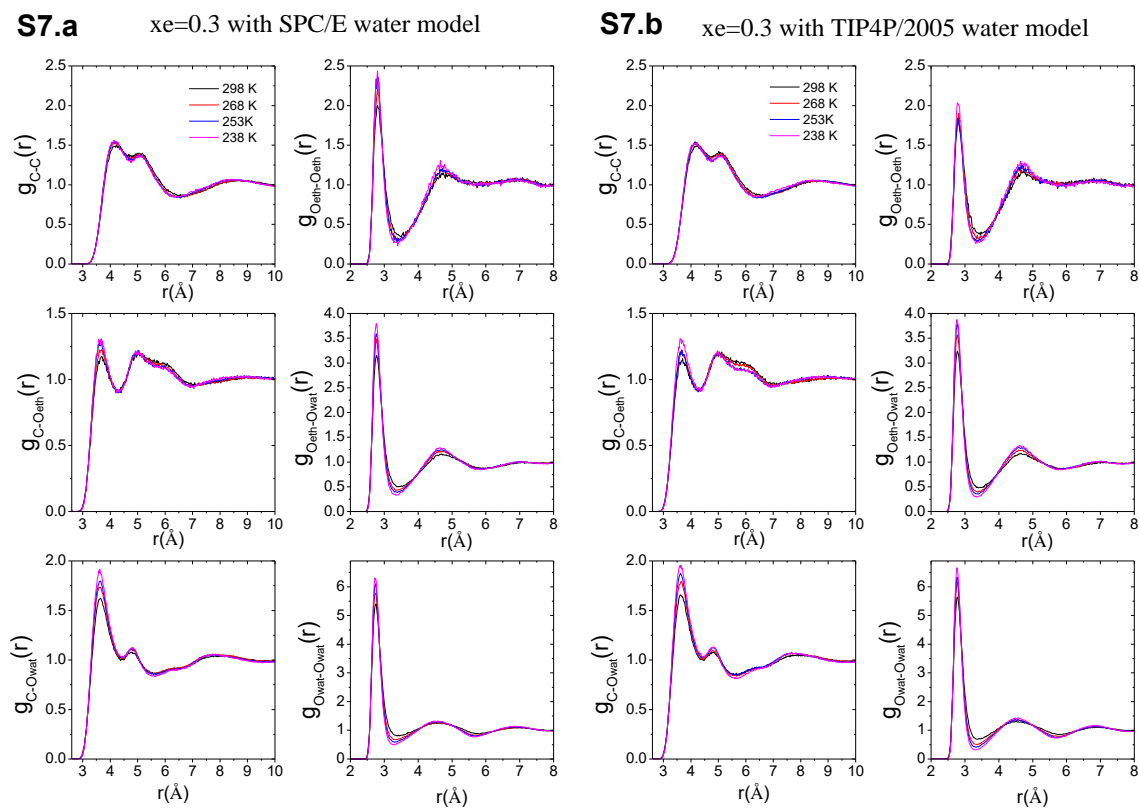

**Figure S8:** H-bond related partial radial distribution functions for the mixture with 10 mol % ethanol, as a function of temperature. a) SPC/E water model; b) TIP4P/2005 water model.

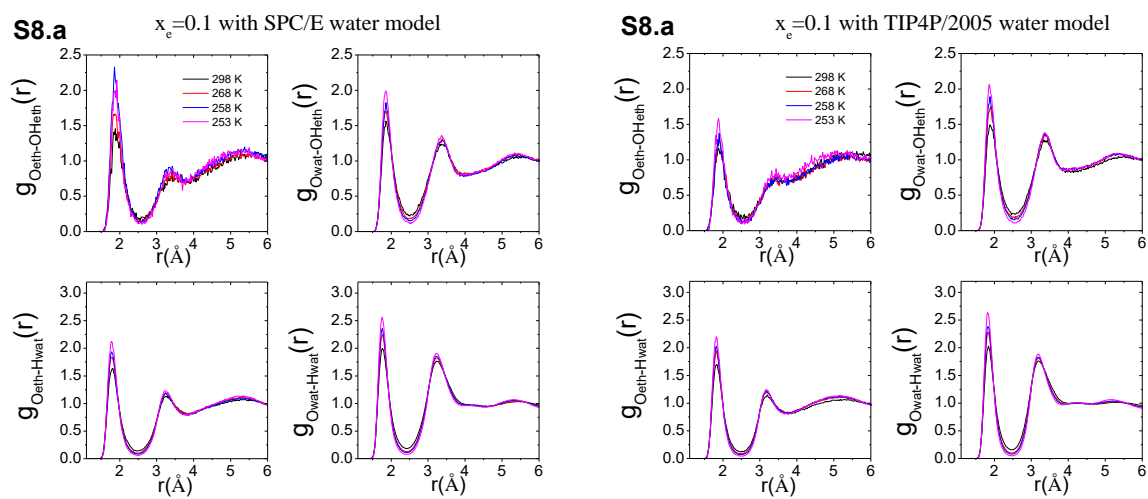

**Figure S9:** H-bond related partial radial distribution functions for the mixture with 30 mol % ethanol, as a function of temperature. a) SPC/E water model; b) TIP4P/2005 water model.

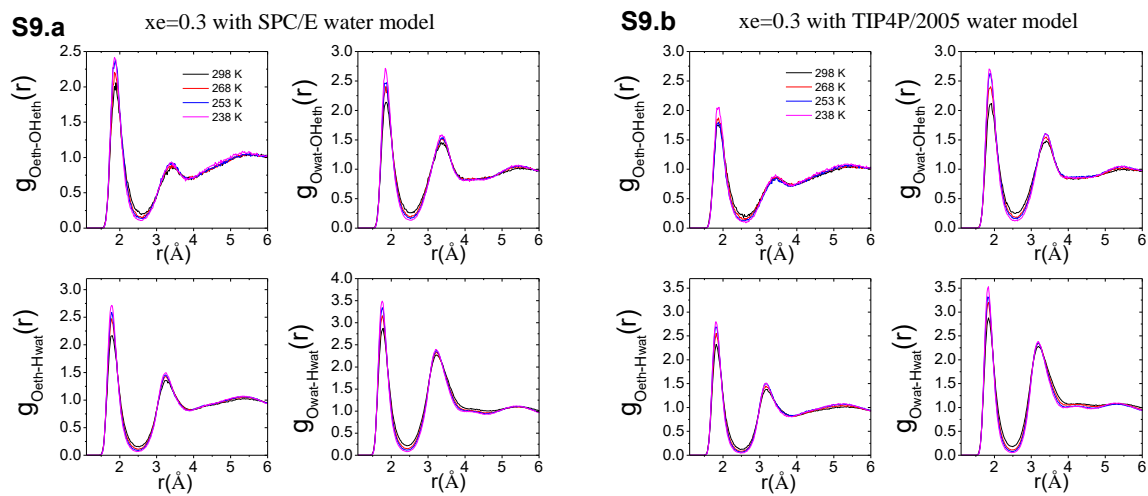

**Table S2**: The characteristic quantities of $g_{\alpha\beta}(r)$ (peak positions, peak heights, coordination numbers calculated up to the first minimum) for the mixture with 20 mol % ethanol.

| $x_e$=0.2 | T(K) | $r_{max}$ | $g(r_{max})$ | $r_{min}$ | $g(r_{min})$ | $n(r_{min})$ |
|---|---|---|---|---|---|---|
| $O_{wat}O_{wat}$ | 298 | 2.74 | 4.57 | 3.46 | 0.79 | 3.74 |
|  | 268 | 2.74 | 4.96 | 3.34 | 0.63 | 3.48 |
|  | 258 | 2.74 | 5.11 | 3.24 | 0.58 | 3.29 |
|  | 253 | 2.74 | 5.21 | 3.32 | 0.56 | 3.45 |
|  | 243 | 2.74 | 5.33 | 3.24 | 0.52 | 3.34 |
|  | 233 | 2.74 | 5.49 | 3.28 | 0.42 | 3.37 |
| $O_{wat}H_{wat}$ | 298 | 1.76 | 2.41 | 2.46 | 0.20 | 0.79 |
|  | 268 | 1.76 | 2.68 | 2.42 | 0.13 | 0.79 |
|  | 258 | 1.76 | 2.78 | 2.42 | 0.12 | 0.78 |
|  | 253 | 1.74 | 2.82 | 2.40 | 0.11 | 0.80 |
|  | 243 | 1.76 | 2.94 | 2.42 | 0.09 | 0.81 |
|  | 233 | 1.76 | 3.05 | 2.46 | 0.07 | 0.81 |
| $O_{eth}O_{wat}$ | 298 | 2.78 | 2.72 | 3.42 | 0.47 | 2.29 |
|  | 268 | 2.78 | 3.02 | 3.34 | 0.42 | 2.29 |
|  | 258 | 2.78 | 3.13 | 3.32 | 0.40 | 2.28 |
|  | 253 | 2.78 | 3.15 | 3.36 | 0.36 | 2.31 |
|  | 243 | 2.78 | 3.29 | 3.34 | 0.33 | 2.29 |
|  | 233 | 2.78 | 3.43 | 3.30 | 0.31 | 2.28 |
| $O_{eth}H_{wat}$ | 298 | 1.80 | 1.90 | 2.50 | 0.15 | 1.32 |
|  | 268 | 1.78 | 2.16 | 2.48 | 0.10 | 1.39 |
|  | 258 | 1.78 | 2.24 | 2.44 | 0.09 | 1.39 |
|  | 253 | 1.78 | 2.25 | 2.48 | 0.08 | 1.39 |
|  | 243 | 1.78 | 2.37 | 2.50 | 0.06 | 1.41 |
|  | 233 | 1.78 | 2.49 | 2.48 | 0.06 | 1.43 |
| $O_{eth}O_{eth}$ | 298 | 2.80 | 1.87 | 3.54 | 0.35 | 0.45 |
|  | 268 | 2.80 | 1.94 | 3.42 | 0.28 | 0.41 |
|  | 258 | 2.80 | 1.99 | 3.48 | 0.28 | 0.42 |
|  | 253 | 2.80 | 2.08 | 3.32 | 0.28 | 0.41 |
|  | 243 | 2.80 | 2.08 | 3.38 | 0.26 | 0.42 |
|  | 233 | 2.80 | 2.19 | 3.36 | 0.23 | 0.41 |
| $O_{eth}OH_{eth}$ | 298 | 1.88 | 1.81 | 2.60 | 0.19 | 0.19 |
|  | 268 | 1.88 | 1.92 | 2.64 | 0.14 | 0.19 |
|  | 258 | 1.88 | 1.97 | 2.64 | 0.13 | 0.18 |
|  | 253 | 1.86 | 2.11 | 2.56 | 0.11 | 0.20 |
|  | 243 | 1.86 | 2.17 | 2.54 | 0.11 | 0.19 |
|  | 233 | 1.86 | 2.23 | 2.58 | 0.08 | 0.20 |
| $C_{eth}O_{wat}$ | 298 | 3.64 | 1.53 | 5.72 | 0.85 | 12.94 |
|  | 268 | 3.60 | 1.72 | 5.64 | 0.87 | 13.54 |
|  | 258 | 3.62 | 1.77 | 5.68 | 0.86 | 13.96 |
|  | 253 | 3.60 | 1.74 | 5.64 | 0.85 | 13.57 |
|  | 243 | 3.66 | 1.80 | 5.62 | 0.83 | 13.68 |
|  | 233 | 3.64 | 1.92 | 5.64 | 0.85 | 14.45 |

**Figure S10:** Ratios of the numbers of ethanol and water molecules that are parts of exactly 'n' cyclic entities, as a function of temperature. Horizontal lines show the expected ratio for the case when the two kinds of molecules behave identically.

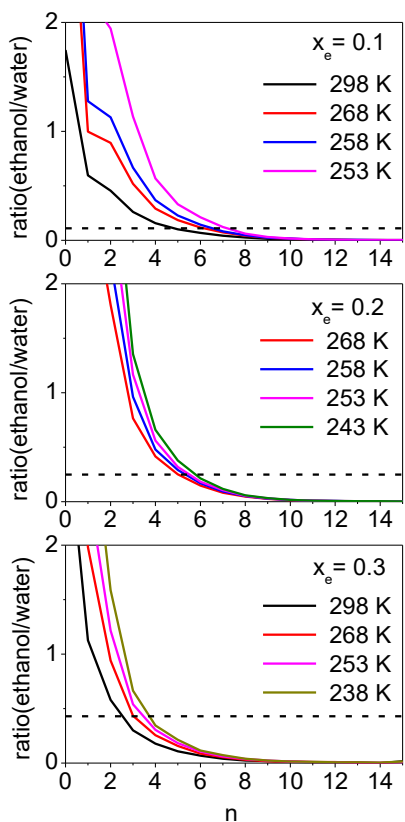